\documentstyle[aps,prl]{revtex}
\def\stackunder#1#2{\mathrel{\mathop{#2}\limits_{#1}}}%
\begin{document}
\title{Interactions of a String Inspired Graviton Field}
\author{Thomas P. Branson$^1$}
\address{Department of Mathematics\\
The University of Iowa\\
Iowa City, IA 52242}
\author{V.G.J.Rodgers$^{2}$ and Takeshi Yasuda$^{3}$}
\address{Department of Physics and Astronomy\\
The University of Iowa\\
Iowa City, IA 52242-1479\\
}

\date{Dec. 1998, revised Dec. 1999}
\maketitle
\footnotetext[1]{thomas-branson@uiowa.edu}
\footnotetext[2]{vincent-rodgers@uiowa.edu (address correspondence)}
\footnotetext[3]{yasuda@hepaxp.physics.uiowa.edu}
\vskip1.0in 
\begin{abstract}
{\sf
We continue to explore the possibility that the graviton in two 
dimensions is related to a quadratic differential that appears in 
the anomalous contribution of the gravitational
effective action for chiral fermions. A higher dimensional
analogue of this field might exist as well.
We improve the defining action for this diffeomorphism tensor 
field and establish a principle for how it interacts 
with other fields and with point particles in any dimension.  
All interactions are related to the action of the diffeomorphism group.
We discuss possible interpretations of this field.  
} 
\vfill
\end{abstract}
To appear in the International Journal of Modern Physics A
\eject
\baselineskip=18pt
\section{Introduction to the Diffeomorphism Tensor}
Recently we introduced a covariant field theory for a rank two tensor
that has its roots in the coadjoint representation of the Virasoro
algebra \cite{emma}. We will call this tensor the
diffeomorphism tensor. In two dimensions one of the components of this rank two
symmetric tensor is the
quadratic differential that appears in the gravitational effective
action for chiral fermions. For this reason we have been compelled 
to explore how 
the field theory for this tensor is related to two
dimensional  gravitation. In \cite{lano2} this analysis was
restricted to the cylinder but a covariant approach was lacking,
while in \cite{emma} a covariant field theory was constructed
emphasizing the use of 
differential expressions which are insensitive to a choice of connection. 
Although that approach gives the correct equations of
motion, it is difficult to relate it to diffeomorphisms.  Furthermore the
conjugate momentum did not transform as an element of the adjoint
representation.  

In this paper we will investigate the relationship between Lie
  derivatives and the Lagrangian for the  diffeomorphism field ${\mbox D}_{\mu
  \nu} $ as well as how the diff field interacts with  matter fields
  such as point particles, fermions and spin one fields.
The action we find that describes the {\it diff tensor} exists in two 
as well as higher dimensions, 
just as Yang-Mills. This suggests that there might be some aspect
of gravitation in two
dimensions that may have a non-trivial link to higher dimensions after all.
Here we write ${\mbox D}_{\mu \nu}$ instead of ${\mbox T}_{\mu \nu}$ 
as has been done in the past \cite{emma,lano2}, so as to avoid
confusion with the energy-momentum tensor.
Our guide to constructing 
a covariant field theory that includes interactions will be the two
dimensional gauge fixed geometric actions and the isotropy algebras on
the coadjoint orbits. From here we can establish a principle that one
can use to identify covariant interactions.  As an
example, we find  the point particle interaction.  
In the presence of a gauge field we find that
the interaction  of the diff field explicitly breaks the gauge
invariance, which can be recovered by introducing a new group valued 
scalar field.

\section{Review: Coadjoint Orbits, Isotropy Groups, and Constraints}
In this section we will  give a short review
of previous work  and then outline  the 
procedure for this work.
The relationship between coadjoint orbits of the Virasoro and
the affine Lie algebras on one hand, and anomalous
contributions to two dimensional effective actions on the other,
has been developed in
\cite{lano2,rai,alek,weigman,lano}. There one is able to construct geometric
actions that correspond to the WZW \cite{witten1} model
and Polyakov \cite{polyakov} gravity.  Physically 
the coadjoint vector (here shown without the central extension), B=(A,D), 
provides the background classical fields A and D that couple to the
bosonized chiral fermions.   The field, A, can be associated with a  
background gauge field of a WZW model \cite{divecchia} while the field
D is typically set to a specific constant to study a fixed Virasoro
orbit. Much of the focus in the literature on two dimensional gravity has
been on the fact that the geometric action of the Virasoro group 
is the Polyakov action.  
Other authors have recognized
the importance of the
coadjoint orbits and  considered {\em model spaces} to treat the different 
coadjoint orbits as elements of a Hilbert space in geometric quantization 
of the Virasoro algebra \cite{Ho}.   Our approach is to carry the
coadjoint representation of the Virasoro algebra beyond conformal
field theory.  The property of the Virasoro orbits that we focus on
is the {\em isotropy algebra} that defines each orbit.  The invariance
of the coadjoint vector due to this subalgebra generates an isotropy
equation that we interpret as a constraint equation among conjugate
variables.  We will explain this point below. Then we simply ask whether
we can construct a covariant field theory reproducing this
constraint for the
background fields associated with the coadjoint vector (A,D).  This
approach allows one to extend our work to a field
theory in higher dimensions.  This does not imply, however, that we now
have a higher dimensional representation of the Virasoro algebra, but rather
that 
there exists a field theory which, when 
dimensionally reduced to a 2D theory, admits
constraints on the fields which agree with the isotropy algebra
associated with the coadjoint representation
of the Virasoro algebra. In what follows we will make these remarks
concrete.

 By looking at the semi-direct product of Virasoro and the
affine Lie algebra on a circle, one gains insight into the
structure of two dimensional gravitation  
through comparison with Yang-Mills. 
We have already seen how the anomalous structures of gravitational and gauge
theories manifest themselves through geometric actions \cite{rai}.   
In \cite{emma,lano2}
the focus was not on the geometric actions, but instead on the field
theories  which describe the dynamics for A and D separately.  We call
these {\em transverse} field theories as
their symplectic structure is transverse to the coadjoint orbits.  
The fields A and D serve as {\em background} 
fields in the geometric
actions and are subjected to constraints  that respect the isotropy
algebra associated with each orbit.  Different orbits correspond
directly to different isotropy algebras.  Our point of view is that
these 
constraint equations are field equations that survive for
a particular choice of static configurations for a gauge field 
${\mbox A}_\mu$
and a diff field ${\mbox D}_{\mu\nu}$.

In \cite{emma} we were interested in the field theory of the diff
field ${\mbox  D}_{\mu \nu}$ alone.
{\em The main interest of this paper is to determine how 
the field ${\mbox D}_{\mu \nu}$ interacts with other fields.}   
As we shall see, the interactions of this rank
two symmetric tensor are  governed by  diffeomorphisms and Lie derivatives.
    
In order to extract any information about the interaction Lagrangian,
we must relate the geometry of the
coadjoint orbits to constraint equations.  Below is a brief 
description of this relationship.
Using the notation of \cite{lano} for the adjoint and coadjoint
representations of the centrally extended
semi-direct product of the Virasoro algebra with an affine Lie algebra,
one can define the algebraic action of a typical adjoint element, say
${\cal F}=\left( \xi \left( \theta \right) ,\Lambda
\left( \theta \right) ,a\right) $, on an arbitrary coadjoint element
${\rm B=%
}\left( {\rm D}\left( \theta \right) ,{\rm A}\left( \theta \right) ,\mu
\right) $
through 
\begin{equation}
{\rm B}_{\cal F}=\left( \xi \left( \theta \right) ,\Lambda \left(
\theta \right) ,a\right) *\left( {\rm D}\left( \theta \right) ,{\rm A}\left(
\theta \right) ,\mu \right) =\left( 
{\rm D}\left( \theta \right) _{{\rm new}%
},{\rm A}\left( \theta \right) _{{\rm new}},0\right) \mbox{.}
\label{coadjoint} 
\end{equation}
Here the components of the new coadjoint elements are
\begin{equation}
{\rm D}\left( \theta \right) _{{\rm new}}=
\;\stackrel{\rm new\;diff\;covector}{%
{}{}{\overbrace{\stackunder{\rm{coordinate\ trans}}{\underbrace{2\xi ^{^{\prime
}}{\rm D}+{\rm D}^{^{\prime }}\xi +\frac{c\mu}{24\pi }
\xi^{\prime \prime \prime }}}-
\stackunder{{\rm gauge\ trans}}{\underbrace{{\rm Tr}\left({\rm A}\Lambda^{\prime }\right) 
}}}}}  
\end{equation}
and
\begin{equation}
{\rm A(\theta )}_{\rm new}=\;\stackrel{\rm new\;gauge\;covector}
{\,\overbrace{%
\stackunder{ {\rm coord\;trans}}{\underbrace{{\rm A}^{\prime }\xi
 +\xi^{\prime }{\rm A}}}
-\stackunder{ {\rm gauge\; trans}}{\,\underbrace{[\Lambda
\,{\rm A}-{\rm A\,}\Lambda ]+k\,\mu \,\Lambda^{\prime }}}}}.
\end{equation}
{}From these transformations one may define the coadjoint orbit of the
the coadjoint element 
${\rm B=}\left( {\rm D}\left( \theta \right) ,
{\rm A}\left( \theta \right) ,\mu
\right) $, as the space of all coadjoint vectors that can be reached
by group transformations on ${\rm B}$.  Each orbit is equipped
with a non-degenerate bilinear two-form that defines a symplectic
structure $\Omega_{{\rm B}}$:  
\[
\Omega _{\rm B}\left( \widetilde{{\rm B}_1},\widetilde{{\rm B}_2}\right)
=\left\langle \;\widetilde{{\rm B}}\;\left| \,\left[ {\rm a}_1{\rm ,a}%
_2\right] \right. \right\rangle ,  
\]
where the coadjoint elements are given by $\widetilde{{\rm B}_i}
{\rm =a}%
_i*\widetilde{{\rm B}}=\delta _{{\rm a}_i}\widetilde{\,{\rm B}}$. 
This symplectic structure is then integrated 
over two-dimensional submanifolds
to extract 
a geometric action that yields the anomalous contribution to the 
effective action of  chiral fermions in the  background fields defined
by  ${\rm B}$.
  
Using 
methods
 from \cite{bal}, one obtains the  action \cite{lano}
\begin{eqnarray}
&S& =\frac 1{2\pi }\stackunder{\mbox{coupling to background
    diffeomorphism field}}
{\underbrace{%
\int {\rm d}\lambda~ {\rm d}\theta~ {\rm d}\tau~ \;{\rm D}\left( \theta \right) \left( \frac{%
\partial _\lambda s}{\partial _\theta s}\partial _\theta \left( \frac{%
\partial _\tau s}{\partial _\theta s}\right) -\frac{\partial _\tau s}{%
\partial _\theta s}\partial _\theta \left( \frac{\partial _\lambda s}{%
\partial _\theta s}\right) \right) }}  \nonumber \\
&+&\frac 1{2\pi }\stackunder{\mbox{coupling to background gauge field}}{\underbrace{%
\int {\rm d}\lambda~ {\rm d}\theta~ {\rm d}\tau~ \;\mbox{Tr}\left( {\rm A}\left( \theta \right)
\left( \frac{\partial _\lambda s}{\partial _\theta s}\partial _\theta \left(
g^{-1}\partial _\tau g\right) -\frac{\partial _\tau s}{\partial _\theta s}%
\partial _\theta \left( g^{-1}\partial _\lambda g\right) +\left[
g^{-1}\partial _\lambda g,g^{-1}\partial _\tau g\right] \right) \right) }} 
\nonumber \\
&-&\frac{\beta c}{48\pi }\stackunder{\mbox{Polyakov gravity}}
{\underbrace{\int
{\rm d}\tau~ {\rm d}\theta~ \;\left( \frac{\partial _\theta ^2s}{\left( \partial _\theta
s\right) ^2}\partial _\tau \partial _\theta s-\frac{\left( \partial _\theta
^2s\right) ^2}{\left( \partial _\theta s\right) ^3}\partial _\tau s\right)
 \label{effective} }}
 \\
&-&\frac{\beta k}{4\pi }\stackunder{\mbox{Wess-Zumino-Witten}}
{\underbrace{\int
{\rm d}\tau~ {\rm d}\theta~ \;{\rm Tr}\left( g^{-1}\partial _\theta gg^{-1}\partial _\tau
g\right) +\frac{\beta k}{4\pi }\int {\rm d}\lambda~ {\rm d}\tau~ {\rm d}\theta~ \;\mbox{Tr}\left(
\left[ g^{-1}\partial _\theta g,g^{-1}\partial _\lambda g\right]
g^{-1}\partial _\tau g\right) \nonumber }},
\end{eqnarray}
where $s(\theta)$ and $g(\theta)$ are diffeomorphism and loop group
elements respectively, representing 
the bosonized fermion degrees of freedom.  Each orbit yields a
distinct system of classical equations of motion.

Since D$(\theta)$ is a function of $\theta$ only, one may integrate by
parts in the 
first line of the above action
and find that the  interaction term for the background field is
\[
S_{I} = \int  {\rm d}\tau~ {\rm d}\theta~ \,  {\rm D} \left( \frac{\partial _\tau    s}{\partial _\theta s}\right).
\]
Thus D appears to be a background field that couples to the induced metric 
$\left( \partial _\tau    s/\partial _\theta s\right)$. 
${\rm D}$ is the ${\rm D}_{\theta \theta}$ 
component of a static
background rank two tensor field.  All other
components of ${\rm D}_{\mu \nu}$   are set to zero as 
they do not couple to dynamical components of the induced
metric.   

In a similar way we know that in the WZW model
the background gauge field interacts with the bosonized fermions 
through the Lagrangian  \cite{divecchia,karabali} 
\begin{equation}
I(g,{\rm A}) = {1\over 4\pi} {\rm Tr}\int {\rm d}\tau~ {\rm d}\theta~ \left({\rm A}_{\theta}\partial_{\tau}g g^{-1} -
  {\rm A}_{\tau}g^{-1}\partial_{\theta}g +{\rm A}_{\theta} g{\rm A}_{\tau}g^{-1} - 
{\rm A}_{\tau}{\rm A}_{\theta}\right).
\end{equation}
By setting $s(\theta) = \theta$ and ${\rm A}_\tau =0$,  Eq.(\ref{effective})
reduces to the WZW model and the interaction Lagrangian 
$$
\frac 1{2\pi }\mbox{Tr} \int {\rm d}\lambda~ {\rm d}\theta~ {\rm d}\tau~ \;
 {\rm A}_\theta 
 \left[g^{-1}\partial _\lambda g,g^{-1}\partial _\tau g\right] 
  = {1\over 4\pi} \mbox{Tr}\int {\rm d}\tau~ {\rm d}\theta~ 
\left({\rm A}_{\theta}\partial_{\tau}g g^{-1} \right).
$$
One sees that the geometric action includes this interaction term
for a particular background  field configuration.  By general
anomaly calculations \cite{balanomaly,divecchia}, we readily identify
the coadjoint vector A with a gauge field.  Just as the
field A$_{\theta}$ is subject to Gauss law constraints,  
the same will hold for the field
D$_{\theta \theta}$; i.e.\ there exist constraints that must be respected 
by the  background field. These constraints manifest themselves in a
subtle but simple way from the point of view of coadjoint orbits,
as they appear through the isotropy algebra of the 
coadjoint element.

The {\bf isotropy group} of a coadjoint element is the group that
leaves that coadjoint element invariant.  Algebraically we have 
for each  coadjoint element,
${\rm B=}\left( {\rm D}\left( \theta \right), 
{\rm A}\left( \theta \right),\mu \right) $;
\begin{equation}
\delta {\rm D}=2\xi_{\rm B} ^{^{\prime}}{\rm D}+{\rm D}^{^{\prime }}\xi_{\rm B} +\frac{c\mu}{24\pi}
\xi_{\rm B}^{\prime \prime \prime }-
{\rm Tr}\left( {\rm A}\Lambda_{\rm B}^{\prime }\right)=0 
\label{codiff} 
\end{equation}
and
\begin{equation}
\delta{\rm A}={\rm A}^{\prime }\xi_{\rm B} +\xi_{\rm B}^{\prime }{\rm A}-[\Lambda_{\rm B} \,{\rm A}-{\rm A\,}\Lambda_{\rm B} ]+k\,\mu \,\Lambda_{\rm B}^{\prime }= 0,
\label{coaffine} 
\end{equation} 
where $\xi_{\rm B} $ and $\Lambda_{\rm B}$ are elements of the isotropy
algebra.   Notice that for a given orbit, these elements are
identified with the identity and do not represent bosonized chiral 
fermion degrees of 
freedom.  {\em We can however
argue that this isotropy algebra originates from constraints
on the canonical variables  A and D and their respective conjugate 
momenta.}  This would illustrate the transverse relationship between the
symplectic structure of an orbit, which 
influences the chiral fermions in the presence of the
background field B, and the symplectic structure that determines
the dynamics for the fields A and D.

\section{Main Results: The Origin of the Diffeomorphism Lagrangian and
  The Interaction Lagrangian}

As mentioned above, the geometric actions associated with the 
coadjoint orbits have a physical interpretation in terms
of chiral fermions coupled to 
gauge and diffeomorphism fields in two space-time dimensions.  In this
 section we discuss the main point of this work, namely 
the determination of the interaction Lagrangian.
We will exploit the coadjoint
representation to aid in determining the interaction Lagrangian for
spin one fields, and borrow from the effective action to extract the
interaction Lagrangian for fermions.  We will then postulate the action for
the  point particle from the point of view of diffeomorphisms.

\subsection{Gauge Transformation Laws and Yang-Mills as a Guide}
We will use the structure of Yang-Mills
theory as a guide to the construction of the interaction
Lagrangian for the diff field.
In Yang-Mills the Gauss law constraint appears as the field equation
arising from varying the Lagrangian with respect to the field 
${\rm  A_0}$ and
evaluating at
${\rm A_0 = 0}$.  In the Yang-Mills case ${\rm A_0}$ can always 
be sent to
zero (i.e., we may reduce to the temporal gauge)
since its conjugate momentum vanishes.  In the case 
of the bosonized fermions, the fields D and
A are the  components of a spin two and spin
one tensor respectively for a particular background field configuration
that couples to the bosonized fermions. Nevertheless we will study the
general covariance for the diff field and show that the isotropy
equations will be analogous to the Gauss law constraints.  

To get an appreciation of the relationship between Gauss' law and the
isotropy equations, 
consider the field A separately from D  and set $\xi = 0$.
Equation~(\ref{coaffine}) represents the residual time-independent
gauge transformations on ${\rm A_\theta}$, where 
${\rm  A}=\left( A_\theta, A_\tau \right)$.
This is a symmetry of the  Cauchy data.

Furthermore by considering  
the transport due to $\xi$ and setting $\Lambda=0$, we see that A 
comes from a tensor of rank one.  That is, the isotropy equation for A is
the space component of a coordinate transformation on ${\rm A}_\mu$,
$${\rm 
\delta_\xi A_\alpha = \xi^\beta \partial_\beta A_\alpha + 
A_\beta \partial_\alpha \xi^\beta.}$$
Under a general coordinate transformation,
we see that the $\xi$ transformation in Equation~(\ref{coaffine}) is the
transformation due to  time independent spatial  translations.  
It is well known that the Yang-Mills action is the requisite action to
covariantly describe ${\rm A_\mu}$ as a dynamical field in any dimension. We have utilized
this fact  in order to understand the pure diffeomorphism
sector, and we  were able to deduce the properties  and the
action for ${\rm D}_{\mu \nu}$ in \cite{emma,lano2}. 
We expect then that we will recover the
isotropy equations as constraint equations for these background
fields.  

With this in mind we  deduce the properties of D, and extend to 
higher dimensions. The analogous constraint equation is 
$$
2\xi ^{^{\prime}}{\rm D}+{\rm D}^{^{\prime }}\xi +\frac{c\mu }{24\pi }\xi^{\prime
  \prime \prime } = 0. 
$$
 Up to the inhomogeneous term, ${\rm D }$ transforms under diffeomorphisms as 
$\delta {\rm D}= 2\xi ^{^{\prime}}{\rm D}+{\rm D}^{^{\prime }}\xi,$ 
corresponding to a rank two tensor, i.e.
$$\xi^a \partial_a 
{\rm D}_{l m} + {\rm D}_{a m} \partial_l \xi^a+{\rm D}_{l a} \partial_m \xi^a
  = \delta_\xi {\rm D}_{l m}. $$

In an analogous way we can use the general coordinate
transformations to produce a ``temporal'' gauge for diffeomorphisms.
In $n$ dimensions one may fix a coordinate system
 so that $\partial_0 {\rm D}_{\mu  0} =0$, leaving the fields 
${\rm  D_{\mu 0}}$ without conjugate momenta.  One can 
solve the $n$ second order equations 
\[
{\partial x^\mu \over \partial x'^0}\big(
{\partial^2 x'^\alpha \over \partial x^\mu \partial x^0} {\partial
  x'^\beta \over \partial x^\rho} {\rm D}_{\alpha \beta} + 
{\partial^2 x'^\beta \over \partial x^\mu \partial x^\rho} {\partial
  x'^\alpha \over \partial x^0} {\rm D}_{\alpha \beta} +
{\partial  x'^\alpha \over \partial x^0}{\partial  x'^\beta \over
  \partial x^\rho}\partial_\mu {\rm D}_{\alpha \beta}\big) = 0,\]
for the coordinate choice. 
As one can see from an infinitesimal coordinate transformation of 
$\partial_0 D_{0 \nu}$, there is a residual ``gauge'' symmetry due to
the time independent coordinate transformations.  
In particular, under the time independent spatial translations ${\rm
  D}_{0 0}$
transforms as a scalar, ${\rm D}_{0 i}$ transforms as a vector and 
${\rm D}_{i j}$ transforms as a rank two tensor. Here the Latin
indices correspond to spatial coordinates.  Separating the time
independent spatial translations from the time independent temporal
transformations on ${\rm D_{\mu 0}}$, we have
\[ 
\delta {\rm D}_{0 i} = {\stackunder{{\rm spatial\, translation}}{\underbrace{
    \xi^j\partial_j {\rm D}_{0 i} + {\rm D}_{0 j} \partial_i \xi^j}}} +
    {\stackunder{{\rm non-tensor\; term}}{\underbrace{ {\rm D}_{0 0} \partial_i \xi^0}}},
\]
while 
\[{\rm \delta {\rm D}_{0 0} = \xi^i\partial_i {\rm D}_{0 0}. }\]  
So we may think of the residual symmetries
as time independent spatial translations 
along with an inhomogeneous transformation for ${\rm D}_{0 i}$.  ${\rm D}_{0 0}$
transforms as a scalar.  We can use $\xi^0$ from the inhomogeneous
term to bring
${\rm D}_{0 1} = 0$.  This leaves another reduced residual symmetry for the
remaining fields as for $i,j \ne 1$, corresponding to $x^0$  and $x^1$
independent transformations,
\[ 
\delta {\rm D}_{0 i} = {\stackunder{x^0 {\rm \,and\,} x^1\ {\rm independent\, translations}}{\underbrace{
    \xi^j\partial_j {\rm D}_{0 i} + {\rm D}_{0 j} \partial_i \xi^j}}} +
    {\stackunder{{\rm non-tensor\, term}}{\underbrace{   \xi^1\partial_1{\rm D}_{0 i}}}}.
\]
We can use $\xi^1$ to bring ${\rm D}_{0 2} = 0$.  

Again this leaves a reduced symmetry and one proceeds to bring all ${\rm D}_{0 i} = 0$.
The ${\rm D_{0 i}}'s$  then serve as Lagrange multipliers for a set of 
constraints.  The constraint hypersurface where ${\rm D}_{0 i}=0$ and
${\rm \partial_0 D_{0 0} = 0}$
is consistent with this analysis.
D then is the one remaining dynamical field
component of a rank two symmetric tensor in two dimensions.

\subsection{Gauss Law Constraints and Field Equations}

{}From the point of view of two-dimensional geometric actions, the 
isotropy group defines the topology of the orbit through 
$(\Omega(G)\otimes{\rm Diff} S^1)/H_{{\rm D,A}}$
where $G$ is the gauge group, $\Omega(G)$ is the loop group 
of $G$, and $H_{{\rm D,A}}$ is the isotropy group of the fields D and A.
The geometric actions then describe the anomalous two-dimensional 
fermionic vacuum in the presence of background gauge and diffeomorphism
fields.  As stated earlier, we are not at all interested
in orbits as they necessarily contain anomalous information. 
Instead we are interested in making A and D dynamical variables
which would certainly move us away from any orbit.  In fact, if we are to
preserve gauge covariance of the initial data, we must guarantee that 
we do not incorporate
gauge variations into the dynamics.  
In field theories the
Gauss  law constraints guarantee that the initial data for the 
dynamical field and its
associated conjugate momentum will not evolve in any (residual) gauge 
directions. The Gauss law constraints are the generators of the 
time independent gauge transformations and spatial translations.
 Thus the dynamical theory of A and D must
be ``transverse'' to the coadjoint orbits.  Since the isotropy condition
 is an equivariant relation between coadjoint elements  and
the adjoint representation, it is precisely the condition that 
defines the Gauss law.  One replaces the coadjoint element that will
serve as the initial data  with the
canonical coordinate and the  adjoint element with the conjugate 
momentum.  This follows since the conjugate momentum transforms like
the adjoint elements.  

In 2D the field 
equations of the ${\rm D}_{0 1}$ component  become constraints on the 
initial data.    
By using the arguments of \cite{emma} one can construct an action such
that the variation with respect to ${\rm D}_{0 1}$ in two
dimensions leads to the isotropy equation for the ${\rm D}_{1 1}$ 
Cauchy data.  

Consider the action
\begin{equation}
S_{\mbox{\tiny diff}} =  \int d^nx \sqrt{g}~
\left( {\rm X}^{\lambda \mu \rho}~{\rm D}^\alpha{}_\rho 
~{\rm X}_{ \mu \lambda \alpha} +2  {\rm X}^{\lambda \mu \rho}~ {\rm D}_{\lambda \alpha}
~{\rm X}^\alpha{}_{\rho \mu}  -\frac{q}4  
{\rm X}^{\alpha \beta}{}_ \beta {} \nabla_\lambda \nabla_\mu{} {\rm X}^{\lambda \mu }{}_\alpha- \frac12  {\rm X}^{\beta \gamma \alpha} 
{\rm X}_{\beta \gamma \alpha} \right). \label{action}
\end{equation}
In the above ${\rm X}^{\mu \nu \rho} = \nabla^\rho {\rm D}^{\mu \nu}$, so we
may write
\[
S_{\mbox{\tiny diff}} =  \int d^nx \sqrt{g}~
\left( (\nabla^\rho {\rm D}^{\lambda \mu})~{\rm D}^\alpha{}_\rho 
~\nabla_\alpha {\rm D}_{\mu \lambda} +2 (\nabla^\rho {\rm D}^{\lambda \mu})~ {\rm D}_{\lambda \alpha}
~\nabla_\mu {\rm D}^\alpha{}_{\rho}  -\frac{q}4  
(\nabla_\alpha\nabla_\beta {\rm D}^{\alpha \beta}){} \nabla_\lambda \nabla_\mu{} 
 {\rm D}^{\lambda \mu }{}- \frac12 (\nabla^\alpha {\rm D}^{\beta \gamma})
\nabla_\alpha {\rm D}_{\beta \gamma } \right). 
\]
This allows us to easily see the three point function, and the fact that $q$
will set a new length scale.
Varying with respect to ${\rm D}_{i 0}$  and setting ${\rm D}_{\nu 0}=0$,
we are led to the equation
\[
{\rm X}^{l m 0} \partial_i {\rm D}^{l m} - \partial_m ({\rm X}^{m l 0} {\rm D}_{l i}) -\partial_l({\rm X}^{m l 0} {\rm D}_{m i}) - 
q{~}\partial_i \partial_l \partial_m {\rm X}^{l m 0}= 0. 
\]
In $1+1$ dimensions, 
this corresponds to the isotropy equation on the coadjoint
orbit specified by D, viz.\\
$\xi {\rm D}' + 2 \xi' {\rm D} + q{~} \xi''' =0$ where $\xi$ takes the
role of ${\rm X}^{1 1 0}$; this is in turn the conjugate momentum for 
${\rm D}={\rm  D}_{1 1}$ after setting ${\rm D}_{i 0}=0$.

\subsection{ Matter Fields Interactions}

The self interaction of the diffeomorphism field suggests that  
interactions with other matter should have the semblance of 
diffeomorphisms.  In \cite{emma} it was suggested that the interaction
Lagrangian be built up from Gauss law constraints that are associated
with the isotropy algebras in one dimension.  
Recall that the interaction 
Lagrangian of the diffeomorphism field had a structure like
\[{\cal L}_{\rm int} = {\rm X^{\lambda \mu \rho}
Y_{\lambda \mu \rho},}\]  where ${\rm X^{\lambda \mu \rho}}$ acts as the 
``covariantized'' conjugate momentum and ${\rm Y_{\lambda \mu \rho}}$ is the
``covariantized'' Lie derivative of the diff field 
${\rm D}_{i j}$.  The construction of the covariant interaction 
Lagrangian 
proceeded in the following stages:
\begin{itemize}
\item  contract the conjugate momentum with a diff variation of the
  matter field;
\item  replace the diff fields $\xi^\alpha$ with ${\rm D}^\alpha{}_0$
 and extend to the Lie derivative;
\item  covariantize the interaction.
\end{itemize}
The interaction Lagrangian is then built up the following stages:
\begin{eqnarray}
{\cal L}_{0^{{\rm th}}~{\rm stage}}&=&{\rm X}^{i j 0}\big(\xi^l 
\partial_l
  {\rm D}_{i j}+{\rm D}_{l j}\partial_i \xi^l +
{\rm D}_{i l}\partial_j \xi^l \big) \rightarrow \nonumber\\
 {\cal L}_{1^{{\rm st}}~{\rm stage}}&=&
{\rm X}^{\lambda \mu 0}\big({\rm D}_0{}^\alpha 
\nabla_\alpha {\rm D}_{\lambda \mu}+{\rm D}_{\alpha \mu}\nabla_\lambda 
{\rm D}_0{}^\alpha +
{\rm D}_{\lambda \alpha}\nabla_\mu {\rm D}_0{}^\alpha \big) \rightarrow
\nonumber\\
 {\cal L}_{\rm int}&=&{\rm X}^{\lambda \mu \rho}\big({\rm D}_\rho{}^\alpha 
\nabla_\alpha  {\rm D}_{\lambda \mu}+{\rm D}_{\alpha \mu}\nabla_\lambda
 {\rm D}_\rho{}^\alpha +
{\rm D}_{\lambda \alpha}\nabla_\mu {\rm D}_\rho{}^\alpha \big). 
\nonumber \end{eqnarray}
With this we have a principle by which we can write the interaction
Lagrangian with other matter.  Notice that even though ${\rm D}_{\mu\nu}$ is a
tensor, one does not simply use tensoriality to find the interactions of 
this field with other matter.  One must require that the one
dimensional theory yield constraint equations that serve as the Gauss
law constraints.   
 
\subsubsection{Fermions}
We will use this principle to build the interaction Lagrangian for 
fermions coupled to the field ${\rm D_{\mu \nu}}$.  
First write the ``covariant'' conjugate momentum for the fermion fields as 
$\sqrt{g}{\bar \Psi} \gamma^\beta$.  
The effect of an infinitesimal diffeomorphism $\xi$ on a spinor $\Psi$
is
$$
\xi^\alpha\nabla_\alpha\Psi
-\frac14\nabla_{[\alpha}\xi_{\beta]}\gamma^\alpha\gamma^\beta\Psi
$$
by, e.g.\ \cite{yks}, 
where $\nabla$ is the spin connection.  Thus
we expect to see
$$
D^\alpha{}_\lambda\nabla_\alpha\Psi
-\frac14\nabla_{[\alpha}D_{\beta]\lambda}\gamma^\alpha\gamma^\beta\Psi.
$$
With this we write the interaction Lagrangian density as
$$
\sqrt{g}\bar\Psi\gamma^\lambda\left(D^\alpha{}_\lambda\nabla_\alpha\Psi
-\frac14\nabla_{[\alpha}D_{\beta]\lambda}\gamma^\alpha\gamma^\beta\Psi
\right).
$$
In order to check to see whether this coupling to fermions is
consistent with two dimensions we can use the geometric action to see
how the bosonized fermions interact with the background field
D.  Recall the 2D result;
\begin{equation}
S = \stackunder{\mbox{fermion-diff}}{\underbrace 
{\int d^2 x~{\rm D}(\theta){ \partial_\tau s \over \partial_\theta s}}}
+ {c \mu  \over 48\pi } \int  \left[
{{\partial^2_{\theta} s}\over{(\partial_{\theta}s)^2}} \partial
_{\tau} \partial_{\theta} s -  {{(\partial^2_{\theta}s)^2
(\partial_{\tau} s)}\over{(\partial_{\theta} s)^3}} \right] {\rm d}\theta~ {\rm d}\tau~.
\end{equation}
In the bosonization of the fermions, 
$({\bar \Psi \gamma^\beta} \partial_\alpha
\Psi)\rightarrow\partial_\tau s/ \partial_\theta s $.  In two
dimensions the coupling to fermions is consistent with our interaction
principle.  This is distinct from the coupling suggested in \cite{emma},
which could also yield the correct two dimensional limit.  However 
that form of the coupling lacks any consistency as to how the diff field
interacts with spin one fields. One sees that the metric has been
replaced with $g_{\mu \nu}+{\rm D}_{\mu \nu}$, suggesting that the
field ${\rm D}_{\mu \nu}$ is playing the role of a classical graviton
field.  We will see that this persists throughout the other
interactions.  

\subsubsection{Spin One Coupling}
The spin one coupling is a good test of this  principle as it is
should have non-trivial contributions to the isotropy equations for
both the A and the D fields.  
Recall that we wish to reproduce the isotropy equation from our field
theory that is reduced to two dimensions. 
The ``covariant'' conjugate momentum for ${\rm A}_\mu$  is
\[ {\rm F}^{\rho \lambda}=
\sqrt{g}\big({\rm \partial^\rho A^\lambda - \partial^\lambda A^\rho
+[A^\rho, A^\lambda] \big)}.\]
The covariantized Lie derivative,
\[{\rm D^\alpha{}_\rho\partial_\alpha
A_\lambda + A_\alpha \partial_\lambda D^\alpha{}_\rho
-\partial_\rho (D^\alpha{}_\lambda A_\alpha)}.\]
is independent of the choice of affine connection.
Therefore the interaction Lagrangian is 
\begin{equation}{\sqrt{g}{\rm F}^{\rho \lambda}
({\rm D}^\alpha{}_\rho\partial_\alpha {\rm A}_\lambda + {\rm A}_\alpha \partial_\lambda
{\rm D}^\alpha_\rho-\partial_\rho({\rm D}^\alpha{}_\lambda
{\rm A}_\alpha))},\label{gauge} \end{equation}
lending no new direct couplings to the metric.  
Notice that when ${\rm D}_{i 0}=0$,   ${\rm A_0}$  has no
conjugate momentum even in the presence of gravity.  
This interaction term, Eq.(\ref{gauge}), is not gauge
invariant.  One may preserve   gauge invariance by introducing a
group valued scalar-field $V$,  transforming under right 
multiplication by a group element $h$ as $V \rightarrow Vh$. 
The interaction Lagrangian is
\begin{equation}{\rm \sqrt{g}F^{\rho \lambda}
(D^\alpha{}_\rho\partial_\alpha {\tilde A}_\lambda + {\tilde A}_\alpha \partial_\lambda
D^\alpha_\rho-\partial_\rho(D^\alpha{}_\lambda
{\tilde A}_\alpha))},\label{gaugenew} \end{equation}
where 
\[ {\rm {\tilde A}}_\mu = {\rm A}_\mu -V^{-1}\partial_\mu V. \]
The $V$ field could have a Lagrangian 
\begin{equation}
{\cal L} = m^2_A \int (V^{-1}\partial_\mu V -
{\rm A}_\mu)(V^{-1}\partial_\nu V -{\rm A}_\nu)(g^{\mu \nu} + D^{\mu
\nu}) d^n x.
\end{equation}
The $V$ field has a coupling constant proportional to the mass
of the gauge fields.  
Variation of the full Lagrangian with respect to ${\rm  D_{1 0}}$,  
followed by evaluation at our background fields with $V=1$, ${\rm A}_0 = 0$,
and ${\rm D}_{0\nu}=0$, 
gives the expected additional  contribution to the constraint
equations predicted by the $2$D constraint equation
\begin{equation}
2{\rm X} ^{^{\prime}}{\rm D}+{\rm D}^{^{\prime }}{\rm X} +\frac{c\mu }{24\pi }{\rm X}^{\prime \prime \prime }-%
{\rm Tr}\left( {\rm A}{{\rm E}}^{\prime }\right) = 0, \label{diff} 
\end{equation}
where X is ${\rm X}^{1 1 0}$ and D is  ${\rm D}_{1 1}$.
Furthermore, variation with respect to ${\rm A}_0$ yields the correct
addition to the constraint via the X dependent terms
\begin{equation}
{\rm A}^{\prime }{\rm X} +{\rm X}^{\prime }{\rm A}-[{\rm E}
\,{\rm A}-{\rm A\,}{\rm E} ]+k\,\mu \,{\rm E}^{\prime } = 0.
\label{gauss}
\end{equation}
These equations are precisely the isotropy equations for 
the background A and D fields, where ${\rm X}$  replaces $  {\xi},$ the 
non-interacting conjugate momentum of D,  and ${\rm E}$ replaces $ {\rm \Lambda},$
the non-interacting conjugate momentum of A, for the background
specified by A and D.      
This one dimensional system was studied both classically and quantum 
mechanically on a cylinder in ref.\cite{lano2}.  The new covariant
picture described here provides a way in which one may reevaluate the Hamiltonian of
the system from
first principles. 

\subsubsection{Point Particle Couplings}
The coupling of ${\rm D_{\mu \nu}}$ to the point particle
helps to illuminate the role of the diff field as the graviton.
Here we proceed in exactly the same way.  We first
identify the conjugate momentum  and multiply it by the diffeomorphism
shift to find that
\[ \int p_i \delta x^i \gamma ~d\tau \rightarrow   \int p_i  D^{i}{}_0 \gamma ~d\tau
\rightarrow  S_{pp}=\int p^\mu p^\nu D_{\mu \nu} {}d\tau.  \]\\
This action, together with the action in  eq.(\ref{action}), suggests a theory with a 
Newtonian potential at low energy for any dimension.  After including
the metric, one has 
\begin{equation}
m { d^2 x^\beta \over d\tau^2}(\delta^\alpha{}_\beta + \mu^2 
{\rm D}_{\beta \mu}g^{\mu \alpha})
+ {d x^\nu\over d\tau} {d x^\lambda\over d\tau} 
\left( \Gamma_{\nu \lambda}{}^\alpha + 
\frac{1}{2 }\mu^2 
g^{\alpha \gamma}(\partial_\gamma {\rm D}_{\nu \lambda} -\partial_\nu {\rm D}_{\gamma \lambda} 
- \partial_\lambda {\rm D}_{\gamma \nu})\right) = 0, \label{eom2}
\end{equation}  
where we have introduced the coupling constant $\mu$.  It is clear
from here that we may interpret the  diff field as a disturbance in the
gravitational field established by the metric, i.e.\ a graviton. 
  The diff field is related to time independent  diffeomorphisms in
 the same way that the gauge fields are related to time independent 
gauge transformations.

\section{Conclusion}
Throughout this paper we have tried to illuminate the role of the
diff field D in gravitation through its interactions with a variety of
matter fields.  We  emphasized the use of isotropy algebras associated
to coadjoint orbits, in order to get
constraints that are related to a field theory. Throughout this work
we have stayed close to Yang-Mills as it has a natural extension to
higher dimensions. 
The way in which D appears in the geometric actions suggests that it
has a classical origin quite distinct from the anomaly.  Different
values of D yield different background symmetries for chiral
fermions, suggesting that there is a semi-classical vacuum structure
that is not anomalous for two dimensional gravity and which is
characterized by different coadjoint orbits.  
Furthermore
the ${\rm D}_{0}{}^{i}$ components act as vector fields generating time
independent coordinate transformations, and guide us to a
working principle for the interaction of gravity with matter.
 This principle can be  carried
over to many other matter fields.  To state this more plainly,
string symmetries have taught us that:
\begin{itemize}
\item two dimensional classical gravity is not trivial, 
\item  but instead 
has a rich structure that is  related to 
diffeomorphisms, and
\item which may  be related
to higher dimensional theories.
\end{itemize}

In two dimensions we find that the diff field behaves as a
``classical'' disturbance of the metric or a classical graviton.
This behavior  is easily seen both
in the case of fermions, and that of a point particle, where the 
metric appears along with the diff field 
as $g_{\mu \nu} + {\rm D}_{\mu \nu}$.   
What is novel here is that apart from scalars and point particles,
the diff field does NOT act as a linear shift of a
fixed metric, but interacts in such a way that the ${\rm D}^i{}_0$
components of the diff field act as elements of the algebra of
diffeomorphisms.   
This yields unexpected results for interactions
with matter; for example the spin one field, but we are comforted by the
fact that these results agree with
the interpretation of the constraints that arise on the coadjoint
orbits in 2D.   We thus use the freedom  due
to generalized coordinate invariance to restrict degrees of freedom on
the diff field (as we discussed using the temporal gauge) 
instead of the metric.   The role of the diff field, as opposed to that
of the metric,
could be that the diff field provides a description of the graviton as a
fluctuation about a metric.  This is distinct from the usual
construction of the graviton as a linearization about a fixed metric.
In our case, the graviton exhibits dynamics that could not have arisen
from linearization about the metric in the Einstein-Hilbert
action. Both the metric and the diff field are
present from this point of view, as the metric is necessary to give a
covariant meaning to the conjugate momentum of the fields and to
define the covariant derivative.  The D field adds further
disturbances to the gravitational field in a way related to
diffeomorphisms, and which is parallel
to the role of the vector potential in gauge theories.   In fact, in
this discussion the vector potential and the diff field are simply two
different components of the three-tuple that defines the centrally
extended coadjoint vector.  The question of geometry
vs.\ diffeomorphisms arises in the theory of gravitation.  
 
The idea of using the Gauss law constraints via the Virasoro algebras
in order to understand gravitation have also been employed
in string field theories  and in lineal gravity
\cite{jackiw}.   In both of these cases the quantum states of the
system carry the adjoint representation of the Virasoro algebra.  
In string field theory one constructs  highest weight states using
negative moded elements of the Virasoro algebra as gauge fixing
conditions, and the other half as generators of gauge transformations.
{}From this procedure one recovers the physical states and ghosts that
are needed for the square of the BRST charge to vanish.   
In lineal gravity one uses a dilaton  
model of gravity to generate a
set of constraints from the commutation relations of the energy-momentum
operators.  This algebra is equivalent to the Virasoro algebra,
 and one requires that the vacuum state be annihilated
by the full Virasoro algebra.  

In our case  we
do not use the adjoint representation, but instead focus on the states
as carriers of the coadjoint representation. The vacuum state of the 
pure diff case will satisfy the operator constraint equation that
arises from classical constraint 
\begin{equation}
{\rm 2X' D + D' X + q X''' = 0.}
\end{equation}
We expect then that the vacuum state will satisfy 
\begin{equation}
{\rm \left( 2 D  {d \over dx} {\delta
  \over \delta D} + {d D \over dx} {\delta \over \delta D} + 
q {d^3 \over dx^3} {\delta \over \delta D} \right) |\Psi> =0.}
\end{equation} 
There is a natural extension of this constraint equation and the
Hamiltonian to higher dimensions giving us direct access to higher
dimensional gravity.  It in not clear how the adjoint representation
can ever have meaning in higher dimensions.  The one feature that may be
dimensionally dependent is that $q$ may be zero in all dimensions
other than two.  After all, the existence of quadratic differentials 
implies having the ability to shift the differentials by an
inhomogeneous term in one space dimension.  (Here the data on the
Cauchy slice acts as the one dimensional quadratic differential.)
However the quantum theory for the higher dimensional theories will
certainly need a quartic term in the propagator, as the three point
function contains two factors of momentum.  This work is being
analyzed presently \cite{yasuda}.  Furthermore, we will extend these ideas to the
super diffeomorphism case to see whether the principle used to construct the
interactions still holds \cite{gates}.

\section{Acknowledgments}
V.G.J.R thanks S.J. Gates, M. Halpern and V.P. Nair for discussion.


\begin{thebibliography}{99}
\bibitem{emma} T. Branson, R.P. Lano, and V.G.J. Rodgers, {\sl Phys. Lett.}
{\bf B412}~{(1997)} 253
\bibitem{lano2} Ralph P. Lano and V.G.J. Rodgers, {\sl
    Nucl. Phys.}~{\bf B437} (1995) 45
\bibitem{kirrilov} A.A. Kirillov, Lect. Notes in Math. 970 (1982) 101,
Springer-Verlag (Berlin)
\bibitem{loop}  A. Pressley and G. Segal, {\sl Loop Groups}, Oxford 
Univ.\ Press (Oxford, 1986)
\bibitem{witten3} E. Witten, {\sl Comm. Math. Phys. }~{\bf 114} (1988) 1
\bibitem{witten1} E. Witten {\sl Comm. Math. Phys.}~{\bf 92} (1984)
  455
\bibitem{divecchia} P. Di Vecchia, B. Durhuus, and J.L. Petersen,
{\sl Phys. Lett.} {\bf B144}~{(1982)} 245
\bibitem{polyakov} A.M. Polyakov, {\sl Mod. Phys. Lett.}~{\bf A2}, No. 11
(1987) 893
\bibitem{rai} B. Rai and V.G.J. Rodgers, {\sl Nucl. Phys.}~{\bf ~B341} 
(1990) 119\\
Gustav W. Delius, Peter van Nieuwenhuizen and V.G.J. Rodgers, 
{\sl Int. J. Mod. Phys.}~{\bf ~A5} (1990), 3943
\bibitem{alek}  A. Yu Alekseev and S.L. Shatashvili,
 {\sl Nucl. Phys.}~{\bf ~B323} (1989) 719
\bibitem{weigman}  P.B. Wiegmann, {\sl Nucl. Phys.}~{\bf ~B323} (1989) 311
\bibitem{yang} C.N. Yang and R.L. Mills, {\sl Phys. Rev.}~{\bf 95} (1954) 631
\bibitem{halpern} K. Bardakci and M.B. Halpern, {\sl Phys. Rev.}~{\bf~ D3} 
(1971) 2493
\bibitem{kac} V.G. Kac, {\sl J. Funct. Anal. Appl.}~{\bf 8} (1974) 68
\bibitem{virasoro} M.A. Virasoro, {\sl Phy. Rev. Lett.}~{\bf ~22} (1969) 37
\bibitem{lano} R.P. Lano and V.G.J. Rodgers, {\sl Mod. Phys. Lett.}~{\bf A7} 
(1992) 1725 \\
R.P. Lano, ``{\sl Application of Coadjoint Orbits to the Loop Group and the 
Diffeomorphism Group of the Circle},'' \\
Master's Thesis, The University of Iowa, May 1994
\bibitem{bal} A.P. Balachandran, G. Marmo, and A. Stern, 
{\sl Nucl. Phys.}~{\bf B162}  (1980) 385;\\
A.P. Balachandran, G. Marmo, B.S. Skagerstam, and A. Stern,
{\sl Nucl. Phys.}~{\bf B164} (1980) 427
 {\sl Phys. Rep. 209} (1991) 129
\bibitem{balanomaly} A.P. ~Balachandran, G.~Marmo, V.P. ~Nair, and
  ~C.G.~Trahern\\
{\sl Phys.\ Rev.\ } {\bf D25}, 2713 (1982)
\bibitem{divecchia} P.~ Di Vecchia,~ B.~Durhuus, and J.L.~Petersen
{\sl Phys.~Lett.} {\bf B144} 245 (1984)
\bibitem{karabali} A.N. Redlich and H.J. Schnitzer. {\sl    Phys.Lett.}~{\bf B167}
 (1986) 315, {\bf B193} (1987) 536\\
 Dimitra Karabali, Q-Han Park, Howard J. Schnitzer,
  Zhu Yang, {\sl Phys.\ Lett.}\ {\bf B216} 307 (1989)
\bibitem{albert} A. Einstein, {\sl Annalen der Physik}, (1919) 49
\bibitem{vgjr} V.G.J. Rodgers, {\sl Phys. Lett.}~{\bf B336} (1994) 343
\bibitem{Ho} Ho-Seong La, Philip Nelson, A.S. Schwarz, {\sl
    Comm. Math. Phys.}~{\bf 134} (1990) 539-554
\bibitem{yks} Y. Kosmann, {\sl Ann. Mat. Pura Appl. (4)}\ {\bf XCI} (1972) 317
\bibitem{jackiw}D. Cangemi, R. Jackiw, B. Zweibach, {\sl Ann. Phys.}
  {\bf B245}, 408 (1996);\\
E. Benedict, R. Jackiw, H.-J. Lee, {\sl Phys. Rev. D} {\bf 54} 6213
(1996)
\bibitem{yasuda} Takeshi Yasuda, work in progress
\bibitem{gates} T.P. Branson,  S.J. Gates and V.G.J. Rodgers, work in progress 
\end{thebibliography}
\end{document}